\documentclass{acm_proc_article-sp}

\usepackage{epsfig,url}

\begin{document}

\title{Users and Assessors in the Context of INEX: Are Relevance Dimensions Relevant?}

\numberofauthors{3}

\author{
%
\alignauthor Jovan Pehcevski\\
       \affaddr{School of CS and IT}\\
       \affaddr{RMIT University}\\
       \affaddr{Melbourne, Australia}\\
       \email{jovanp@cs.rmit.edu.au}
\alignauthor James A. Thom\\
       \affaddr{School of CS and IT}\\
       \affaddr{RMIT University}\\
       \affaddr{Melbourne, Australia}\\
       \email{jat@cs.rmit.edu.au}
\alignauthor Anne-Marie Vercoustre\\
       \affaddr{AxIS research group}\\
       \affaddr{INRIA}\\
       \affaddr{Rocquencourt, France}\\
       \email{anne-marie.vercoustre@inria.fr}
}

\maketitle

\begin{abstract}

{\em
The main aspects of XML retrieval
are identified by analysing and comparing the following two behaviours:
the behaviour of the assessor when judging the relevance of returned document components; and
the behaviour of users when interacting with components of XML documents.
We argue that the two INEX relevance dimensions, Exhaustivity and Specificity, 
are not orthogonal dimensions; indeed, an empirical analysis of each dimension
reveals that the grades of the two dimensions 
are correlated to each other.
By analysing the level of agreement between the assessor and the users,
we aim at identifying the best units of retrieval.
The results of our analysis show that the highest level of agreement 
is on highly relevant and on non-relevant document components, 
suggesting that only the end points of the INEX 10-point relevance scale 
are perceived in the same way by both the assessor and the users.
We propose a new definition of relevance for XML retrieval and argue that its corresponding relevance scale
would be a better choice for INEX.\/}

\end{abstract}


\section{Introduction}

The INitiative for the Evaluation of XML retrieval\footnote{{\tt http://inex.is.informatik.uni-duisburg.de/2005/}} (INEX) 
is a coordinated effort that
promotes evaluation procedures for content-oriented XML retrieval.
In order to evaluate XML retrieval effectiveness,
the concept of \emph{relevance} needs to be clearly defined. 
There are two \emph{relevance dimensions} used by INEX,
\emph{Exhaustivity} and \emph{Specificity},
which measure the extent to which a given information unit \emph{covers} and is \emph{focused on} an information need, respectively~\cite{Benjamin-Lalmas}.
In this paper we provide a detailed empirical analysis of the two INEX relevance dimensions.
More specifically, we investigate what the experience of both the assessor and the users suggests on 
how relevance should be defined and measured in the context of XML retrieval.

The INEX test collection
consists of three parts: an XML document collection, a set of topics required to search for information stored in this collection, and a set of
relevance assessments that correspond to these topics~\cite{INEX04-Overview}. 
The XML document collection comprises 12,107 IEEE Computer Society
articles published in the period between 1997-2002, with approximately 500MB of data.
To search for information stored in this collection,
two types of topics are explored in INEX: 
Content-Only (CO) topics and Content-And-Structure (CAS) topics.
CO topics do not refer to the existing document structure,
whereas CAS topics
enforce restrictions on the document structure and
explicitly specify the target element. In this paper, we focus on the CO topics 
to analyse the behaviour of the assessor and the users in the context of INEX.

Tombros et al.~\cite{Tombros2005} demonstrate that,
while assessing relevance of retrieved pages on the Web,
 the context determined by a task type has an effect on the user behaviour. 
A similar effect is likely to be expected when users assess the relevance of XML document components 
(rather than of whole documents, such as Web pages)~\cite{Tombros-LNCS2005}. 
The CO topics used in this study are thus selected such that they correspond to different types of tasks, 
or different \emph{topic categories}: a \emph{Background} category and a 
 \emph{Comparison} category.

Since 2002, a new set of topics has been introduced
and assessed by INEX participants each year.
Analysing the behaviour of assessors when judging the relevance of returned document components
may provide insight into the possible trends
within the relevance judgements. 
Such studies have been done for both the 
INEX 2002~\cite{Kazai-ECIR04} and the INEX 2003~\cite{Benjamin-Lalmas} test collections.
We have recently also analysed the relevance judgements of the INEX 2004 topics, 
where we aimed at understanding what assessors consider to be the most useful answers~\cite{Pehcevski-LNCS2005}. 

There is growing interest among the research community in studying
the user behaviour in the context of XML retrieval;
however, little work has been done in the field so far.
The most notable is the work done by Finesilver and Reid~\cite{Finesilver-ECIR03},
where a small-scale experimental study is designed to investigate the 
information-seeking behaviour of users in the context of structured documents. 
Recently, an Interactive track was established at INEX 2004 to investigate the retrieval behaviour of users
when components of XML documents -- estimated as likely to be relevant by an XML retrieval system -- 
are presented as answers~\cite{Tombros-LNCS2005}. 
Ten of the 43 active research groups in INEX 2004 were also involved in the Interactive track, and
each group was required to provide a minimum of eight users to interact with the retrieval system. 
The analysis of the user behaviour in this paper is based on the user judgements provided by these groups.

When judging the relevance of a document component, two relevance dimensions -- \emph{Exhaustivity} and \emph{Specificity} -- are used by INEX. 
Each dimension uses four grades of relevance. 
To assign a relevance score to a document component, the grades from each dimension are combined into a single 10-point relevance scale. 
However, the latter choice of combining the grades poses the following question: 
\emph{is the 10-point relevance scale well perceived by users}?

Due to hierarchical relationships between the XML document components, 
an XML retrieval system may often return components with varying granularity.
The problem that often arises in this retrieval scenario is 
the one of distinguishing the \emph{appropriate level of retrieval granularity}.
This problem, which is often referred to as the \emph{overlap problem}, remains
an open research problem in the field of XML retrieval.
Indeed, it has been shown that it is not only a retrieval problem~\cite{Pehcevski-LNCS2005,Pehcevski-Kluwer2005}, 
but also a serious evaluation problem~\cite{Kazai/etal:04}. 
This then raises the question: \emph{is retrieving overlapping document components what users really want}? 

In this work, we aim at finding answers to the above research questions. 
We show that the overlap problem is handled differently
 by the assessor and the users,
and that the two INEX relevance dimensions are 
perceived as one.
We propose a new definition of relevance for INEX and argue that 
its corresponding relevance scale would bring a better value for the XML retrieval evaluation.

The remainder of the paper is organised as follows.
In Section 2 we provide an overview of the methodology used in this study. 
The concept of \emph{relevance} in information retrieval is thoroughly discussed in Section 3,
where we particularly focus on how the INEX definition of relevance fits in the unified relevance framework.
We study the behaviour of the assessor and the users in Sections 4 and 5,
when two categories of retrieval topics are considered, respectively. 
Our new definition of relevance is described in Section 6.
We conclude in Section 7 with a brief discussion of our findings.

\section{Methodology}

In this section, we provide a detailed overview of the methodology used in this study. 
More precisely, 
we describe the type and the number of participants involved;
the choice of the two categories of topics used;
and the way the data -- reflecting the observed behaviour of participants -- was collected.
The data reflecting the observed behaviour, as analysed in this study, 
was collected from well-established INEX activities, which are also explained in separate studies. 
For instance, for a particular CO topic 
we use the relevance judgements obtained from the interactive online assessment system~\cite{Benjamin-Lalmas} 
to analyse the behaviour of the assessor.
Similarly, for the same topic we use the data collected for the purposes of the INEX 2004 Interactive track~\cite{Tombros-LNCS2005}
to analyse the retrieval behaviour of users.
We actively participated in both INEX activities.

\subsection{Participants}

Two types of participants are used in this study: assessors and users.
In general, both can be regarded as users; 
however, it is often necessary to distinguish between them, since their purpose 
in the XML retrieval task is quite different.

\subsubsection*{Assessors}

Every year since 2002 when INEX started, each participant is asked to submit at least one retrieval topic (query). If 
a topic is accepted, the same participant is (usually) required to assess the relevance of the retrieved document components.
 The assessor can, therefore, be seen as an entity that provides the ground-truth 
for a particular retrieval topic. There is usually one assessor per topic, 
although for the purpose of checking whether the relevance judgements were done in a consistent manner, 
two or more assessors may be assigned to a given topic~\cite{Benjamin-Lalmas}. 
In this study we analyse the relevance assessments provided by one assessor per topic.

\subsubsection*{Users}

A total of 88 users were employed for the purposes of the Interactive track at INEX 2004, 
with an average age of 29 years~\cite{Tombros-LNCS2005}. Although most of the users had a substantial level of 
experience in Web or other related searches, it was expected that very few (if any) were experienced in interacting with XML document components.
For this purpose, users were given the same (or rather, slightly modified) retrieval topics as the ones proposed and judged by the assessors.
Analysing the data collected from the user interaction
may thus indicate how well an XML retrieval system succeeds in satisfying users' information needs.
Our analysis in this study is based on the user judgements provided by roughly 50 users per topic.

\subsection{Retrieval Topics}

\begin{figure}[tb]
\begin{picture}(0,0)
\thicklines
\put(0,0){\line(1,0){245}}
\end{picture}
\begin{small}
\begin{verbatim}
Topic B1 (INEX 2004 CO topic 192):

You are writing a large article discussing virtual reality
(VR) applications and you need to discuss their negative 
side effects. What you want to know is the symptoms 
associated with cybersickness, the amount of users who get 
them, and the VR situations where they occur. You are not 
interested in the use of VR in therapeutic treatments 
unless they discuss VR side effects.
\end{verbatim}
\end{small}
\begin{picture}(0,0)
\thicklines
\put(0,0){\line(1,0){245}}
\end{picture}
\caption{A \emph{Background} topic example.}
\label{TopicB1}
\end{figure}

To make users better understand the objectives of the retrieval task,
the CO topics were reformulated as simulated work task situations~\cite{Tombros-LNCS2005}. 
A simulated work task situation requires users to interact with the retrieval system, 
which in turn -- by allowing users to formulate as many queries as needed -- 
results in different individual interpretations of the information need~\cite{Borlund2003}.
Thus, the reformulated CO topics not only describe \emph{what} the information need represents, but also \emph{why}
users need to satisfy this need, and what is the \emph{context} where the information need arises.

The CO topics used in the INEX Interactive track are divided in two task categories: a \emph{Background} category and a 
 \emph{Comparison} category. Topics that belong to the \emph{Background} category seek to find as much general information about 
the area of interest as possible. 
Two retrieval topics were used in this category, B1 and B2,
which are based on the INEX 2004 CO topics 192 and 180, respectively~\cite{Tombros-LNCS2005}.
Figure~\ref{TopicB1} shows Topic B1, which is the \emph{Background} topic used in this study.
Topics that belong to the \emph{Comparison} category seek to find similarities or differences between at least two items discussed in the topic.
Two retrieval topics were used in this category, C1 and C2, which
are respectively based on the INEX 2004 CO topics 188 and 198~\cite{Tombros-LNCS2005}.
Figure~\ref{TopicC2} shows Topic C2, which is the \emph{Comparison} topic used in this study.

The motivation of using topics B1 and C2 in our study 
comes from the fact that both of these topics have corresponding relevance judgements available, 
and that data from roughly 50 users was collected for each of these topics. 
In contrast, no relevance judgements are available for
topic B2, while data from around 18 users was collected for
each of the topics B2 and C1.
Previous work has also shown that XML retrieval systems exhibit varying behaviour 
when their performance is evaluated against different CO topic categories~\cite{Hatano/etal:04,Pehcevski-Kluwer2005}.
It is then reasonable to expect that the level of agreement between the assessor and the users, 
which concerns the choice of the best units of retrieval, may 
depend on the topic category.
Thus, in our forthcoming analysis of the retrieval behaviour,
we clearly distinguish between topics B1 and C2.

\subsection{Collecting the Data}

\begin{figure}[tb]
\begin{picture}(0,0)
\thicklines
\put(0,0){\line(1,0){245}}
\end{picture}
\begin{small}
\begin{verbatim}
Topic C2 (INEX 2004 CO topic 198):

You are working on a project to develop a next generation 
version of a software system. You are trying to decide on 
the benefits and problems of implementation in a number
of programming languages, but particularly Java and Python. 
You would like a good comparison of these for application 
development. You would like to see comparisons of Python 
and Java for developing large applications. You want to 
see articles, or parts of articles, that discuss the 
positive and negative aspects of the languages. Things 
that discuss either language with respect to application 
development may be also partially useful to you. Ideally, 
you would be looking for items that are discussing both
efficiency of development and efficiency of execution 
time for applications.
\end{verbatim}
\end{small}
\begin{picture}(0,0)
\thicklines
\put(0,0){\line(1,0){245}}
\end{picture}
\caption{A \emph{Comparison} topic example.}
\label{TopicC2}
\end{figure}

Different means were used to collect the data from the assessor and the users, and different time restrictions 
were put in place in both cases. 

In the case of the assessor, an interactive online assessment system is used to 
collect the judgements for a particular topic~\cite{Benjamin-Lalmas}.
This is a well-established method used in INEX, 
where the assessment system implements some rules to ensure that 
the collected relevance judgements are as exhaustive and as consistent as possible. 
On average it takes one week for the assessor 
to judge all the retrieved elements for a particular topic.
The relevance judgements are then stored in an \emph{XML assessment file} where, 
for each XML document retrieved by participant systems, 
the judged elements are kept in document order.
We use two assessment files, one for each topic B1 and C2, 
to analyse the relevance judgements made by assessors.

For users, a system based on HyREX~\cite{Hyrex} 
is used to collect the user judgements and to log their activities. 
Tombros et al.~\cite{Tombros-LNCS2005} explain the process of user interaction with the HyREX system in detail.
Users are able to choose between two retrieval topics for each topic category, 
for which they are required to find as much information as possible 
for completing the search task.
A time limit of 30 minutes is given to each user.
The data obtained from the user interaction is stored in corresponding log files.
For each user, we create an assessment file  
that follows the same structure as the assessor's assessment file.
We use these files to analyse the judgements made by users for each of the topics
B1 and C2.

An important point to note is that HyREX uses the concept of ``index objects''~\cite{Hyrex} 
to limit the level of retrieval granularity that will be returned to users. 
This means that users were able to make judgements for only four (out of 192) element names.
These names are \verb+article+, \verb+sec+, \verb+ss1+, and \verb+ss2+, which correspond to
full article and to section and subsection elements of varying nesting levels, respectively. 
Although this may be seen as a limitation of the HyREX system, 
the obtained element granularity is nevertheless sufficient for the purpose of our analysis. 
To be consistent in our comparison of the observed behaviour between the assessor and the users, 
all element names different from these four were also removed from the 
two files containing assessors' judgements. 
If an element has been judged more than once, either by a user or an assessor, only the last 
relevance judgement is stored in the assessment files.

\subsection{Measuring Overlap}

When collecting assessor or user judgements for a particular topic,
we also measure the level of overlap between the judged elements.
There are \emph{at least} two ways by which the overlap can be measured:
\begin{itemize}
\item \emph{set-based overlap}, which for a \emph{set} of returned elements
measures the percentage of elements for which there exists another element 
that \emph{fully contains} them; and 
\item \emph{list-based overlap}, which takes into account the order of processing of returned elements,
and measures the percentage of elements for which there exists another
element \emph{higher in the list} that fully contains them.
\end{itemize}

Consider the following set of returned elements:

\begin{small}
\begin{enumerate}
\item \verb+/article[1]/sec[1]+
\item \verb+/article[1]/sec[1]/ss1[1]+
\item \verb+/article[1]/sec[1]/ss1[1]/ss2[1]+
\item \verb+/article[1]/sec[2]/ss1[1]+
\item \verb+/article[1]/sec[2]+
\end{enumerate}
\end{small}

Let us assume that the elements are returned in the above order, and that all the elements belong to one XML document.
The set-based overlap in this case would be 60\%, because three (out of five) elements
in this set are fully contained by other element in the set (the three elements are the ones belonging to ranks 2, 3 and 4). 
The list-based overlap, however, would be 40\%, because there are only two
elements for which there exists another
element higher in the list that fully contains them (the two elements that belong to ranks 2 and 3).

In this study we use the set-based overlap, as defined above, to measure the overlap 
between the judged elements.
However, unlike in the assessor's case where the relevance judgements were obtained from only one assessor, 
the user judgements for a given topic were obtained from more than one user.
To deal with this issue in a consistent manner,
in users' case we measure the overlap \emph{separately for each user}, 
and take the average to represent the resulting set-based overlap.

\section{Relevance: Definitions and \\Dimensions}

It is a commonly held view that \emph{relevance} is one of the most important concepts 
for the fields of documentation, information science, and information retrieval~\cite{Mizzaro,Saracevic}. 
Indeed, 
the main purpose of a retrieval system is to retrieve units of information
estimated as \emph{likely to be relevant} to an information need, as represented by a query.
To build and evaluate effective information retrieval systems, 
the concept of relevance needs to be clearly defined and formalised. 

Mizzaro~\cite{Mizzaro} provides an overview of different definitions of relevance.
These are also conveniently summarised by Lavrenko~\cite{Lavrenko-PhD}.
In general, there is a system-oriented, a user-oriented, and a logical definition of relevance.
However, there are also other definitions of relevance, which relate to its nature and the notion of dependence.
With respect to its nature, there is a binary or non-binary (graded) relevance.
With respect to whether the relevance of a retrieved unit is dependent or not on
any other unit already inspected by the user, there is a dependent or independent relevance.
In the case of the former, the relevance is often distinguished either 
as a relevance conditional to a set of relevant retrieved units,
or as a novel relevance, or as an aspect relevance. 

In the following we provide an overview of several definitions of relevance,
including the INEX relevance definition. 
We then describe a notable attempt to construct a unified definition of relevance~\cite{Mizzaro}.

\subsection{System-oriented Relevance Definition}

The system-oriented definition
provides a binary relation between a unit of information (a document or a document component) and a user request (a query).
To model this relation, both the unit of information and the user request are represented by a set of terms, 
reflecting the contents of the unit and the interest of the user, respectively. 
In this case, relevance is simply defined by the level of semantic overlap between the two representations; 
the more similar these representations are, the more likely 
the information unit is relevant to the user request.
According to this definition, relevance 
is not dependent on any factors other than the two representations above.
More precisely, it depends neither on the user who issued the request (or on the user information need, for that matter), 
nor on any other information units (regardless of whether they have been previously considered to be relevant or not), 
nor on any other requests to which the unit of information may or may not be relevant.

\subsection{Novel Relevance}

Novel relevance deals with the impact of retrieving redundant information units on user's perception of relevance. 
For example, if a system retrieves two near-duplicate information units, which may both be relevant to a request, 
the user will very likely not be interested in reading both of them, since once the first one is read, the second 
becomes entirely redundant. 
Carbonell and Goldstein proposed the concept of \emph{Maximal Marginal Relevance}~\cite{MMR},
which attempts to provide a balance between the relevance of a document to a query, and the redundancy of that document 
with respect to all the other documents previously inspected by the user. 
An interesting approach that may be seen as an extension of the above work was proposed by Allan et al.~\cite{Allan-aspect}.
Their work attempts to address redundancy on a sub-document level and
is based on the following idea: even if a document is considered to be mostly redundant by a user, 
it may still contain a small amount of novel information
(which is, for example, often the case in news reporting). 
Therefore, they independently evaluate the performance of an information retrieval system
with respect to two separate definitions of relevance: a topical relevance and a novel relevance.
We believe that this (or a similar) approach is particularly attractive for the field of XML retrieval,
where systems tend to retrieve mutually overlapping (and thus redundant) information units.
Some aspects of novel relevance are investigated in detail by the TREC Novelty track~\cite{TREC2003-novelty}.

\subsection{Aspect Relevance}

A user request often represents a complex information need that 
may comprise smaller (and possibly independent) parts, often called \emph{aspects}. 
The goal of an information retrieval system is then to retrieve information units that cover as many aspects of
the information need as possible. 
In this context, \emph{aspect relevance} is defined as topical relevance of the retrieved unit to 
a particular aspect of the information need, whereas
\emph{aspect coverage} is defined as the number of aspects for which relevant retrieved units exist.
Zhai~\cite{Zhai-PhD} describes a formal approach to modelling aspect relevance. 
INEX uses a somewhat modified definition of aspect relevance, which will be discussed in more detail below.

\subsection{The INEX Relevance Definition}

From 2003 in INEX, the relevance of an information unit (a document or a document component) to a request (a query) is described by two 
dimensions: \emph{Exhaustivity}, which represents topical relevance 
that models the extent to which the information unit discusses aspects of the information need
represented by the request, and 
\emph{Specificity}, which also represents topical relevance,
but models the extent to which the information unit focuses on aspects of the information need.
For example, an information unit may be highly exhaustive to a user request (since it discusses most or all the aspects of the information need),
but only marginally specific (since it also focuses on aspects other than those concerning the information need).
Conversely, an information unit may be highly specific to a user request (since there is no non-relevant information and 
it only focuses on aspects concerning the information need),
but it may be marginally exhaustive (since it discusses only a few aspects of the information need).

In traditional information retrieval, 
a binary relevance scale is often used to assess the relevance of an information unit (usually a whole document) 
to a user request\footnote{Recent Robust and Web tracks in TREC, however, use a non-binary relevance scale for evaluation.}.
The relevance value of the information unit is restricted to either zero
(when the unit is not relevant to the request) or one (when the unit is relevant to the request). 
INEX, however, adopts a four-graded relevance scale for each of the relevance dimensions, 
such that the relevance of an information unit to a request ranges from none, to marginally, to fairly, or to highly exhaustive or specific,
respectively. To identify \emph{relevant} units of information, that is, units of information that are both 
exhaustive and specific to a user request, a combination of the grades from each of the two relevance dimensions is used. 
These relevant units are then, according to INEX, ``the most appropriate units of information to return as an answer to the query''~\cite{Benjamin-Lalmas}.
Table~\ref{10-point} shows the combination of the grades from each of the two relevance dimensions, which represents the 10-point relevance scale used by INEX.

\begin{table}[tp]
\begin{center}
\begin{tabular}{l c c c c c}
\hline
 & & \multicolumn{4}{c}{\bf{Exhaustive}}  \\
 \cline{3-6}
\multicolumn{1}{c}{\bf{Specific}} & & \emph{Highly} & \emph{Fairly} & \emph{Marginally} & \emph{None} \\
\hline \hline
\emph{Highly} & & \bf{E3S3} & \bf{E2S3} & \bf{E1S3} & \bf{E0S0} \\ 
\emph{Fairly} & & \bf{E3S2} & \bf{E2S2} & \bf{E1S2} & \bf{E0S0} \\ 
\emph{Marginally} & & \bf{E3S1} & \bf{E2S1} & \bf{E1S1} & \bf{E0S0} \\ 
\emph{None} & & \bf{E0S0} & \bf{E0S0} & \bf{E0S0} & \bf{E0S0} \\ 
\hline \hline
\end{tabular}
\end{center}
\caption{The 10-point relevance scale, as adopted by INEX. Each point of the relevance scale 
combines a particular grade from the Exhaustivity dimension with a corresponding grade from the Specificity dimension.}
\label{10-point}
\end{table}

The two relevance dimensions, \emph{Exhaustivity} and \emph{Specificity}, are not completely independent. 
An information unit that is not exhaustive is at the same time not specific to the request (and vice versa), which restricts the 
space of combining the grades of the two dimensions to ten possible values. In the remainder of the paper, a relevance value of
an information unit to a request will be denoted as \verb+EeSs+, where \verb+E+ represents \emph{Exhaustivity}, 
\verb+S+ represents \emph{Specificity}, and \verb+e+ and \verb+s+ represent integer numbers between zero and three.
For example, \verb+E1S3+ represents an information unit that is marginally exhaustive and highly specific to a request.
An information unit is considered relevant only if both \verb+e+ and \verb+s+ are greater than zero. 
The relevance value \verb+E0S0+ therefore denotes a non-relevant information unit, whereas 
the value \verb+E3S3+ denotes a highly relevant information unit.

\subsubsection*{Comparison with Aspect Relevance}

A strong parallel may be drawn between \emph{Exhaustivity} and \emph{Specificity}, the two INEX relevance dimensions, 
with \emph{aspect coverage} and \emph{aspect relevance}. 
\emph{Exhaustivity} maps the aspect coverage to a four-point relevance scale, 
from \verb+E0+ being ``the XML element does not discuss the query at all''~\cite{Benjamin-Lalmas}, to \verb+E3+ being
``the XML element discusses most or all aspects of the query''~\cite{Benjamin-Lalmas}.
\emph{Specificity}, on the other hand, is almost identical to aspect relevance.

\subsection{Unified Relevance Definition}

A notable attempt to construct a unified definition of relevance is given by Mizzaro~\cite{Mizzaro}.
He formalises a framework capable of modelling various definitions of relevance by embedding it in a four-dimensional space.

The first dimension deals with the type of entities for which the relevance is defined. It can take one of the 
following three values: \emph{Document}, \emph{Surrogate}, or \emph{Information}.
\emph{Document} refers to the information unit a user will obtain as a result of their search; this may represent 
a full-text document, an image, video, or, in the case of XML retrieval, a document component.
\emph{Surrogate} refers to a form of representation of \emph{Document}; this may be of a set of terms, 
bibliographic data, or a condensed abstract of the information unit. 
The third value, \emph{Information}, refers to a rather abstract concept, which 
depends on the type and amount of information 
the user receives while reading or consuming the contents of the returned unit of information.

The second dimension relates to the level at which the user request is dealt with. There are four possible levels:
\emph{Problem}, \emph{Information need}, \emph{Request}, or \emph{Query}. 
The \emph{Problem} (also referred to as Real Information Need -- RIN~\cite{Lavrenko-PhD}) 
relates to the actual problem that a user is faced with, and for which information is needed to help solve it.
The user may not be fully aware of the actual problem; instead, in their minds they perceive 
it by forming a mental image. This mental image in fact represents the \emph{Information need}
(also referred to as Perceived Information Need -- PIN~\cite{Lavrenko-PhD}).
\emph{Request} is a way of communicating the \emph{Information need} to others by specifying it in a natural language.
For the \emph{Request} to be recognised by a retrieval system, it needs to be represented by a \emph{Query}.
The \emph{Query} usually consists of a set of terms, optionally including phrases or logical query operators.

Relevance can then simply be seen as a combination of 
any of the entities from the two dimensions above;
that is, it can be seen as a combination of any of the \emph{values} from the first dimension 
with any of the \emph{levels} from the second dimension. 
Indeed, phrases such as ``relevance of a \emph{Surrogate} to a \emph{Query}'' or 
``relevance of a \emph{Document} to a \emph{Request}'' are often used.
Mizzaro, however, argues that this relevance space does not actually represent the space of all possible relevances.
Rather, there is also a third dimension that specifies the nature of the relationship between the two dimensions.
The components of this third dimension are \emph{Topic}, \emph{Task}, \emph{Context}, or any combination of the three.
The \emph{Topic} (or topical relevance~\cite{Lavrenko-PhD}) specifies how similar the two entities are to 
user's area of interest. For example, if the user is interested 
in finding information about the overlap problem in XML retrieval,
the topical relevance will represent the level of similarity of the retrieved unit to the query with respect to that particular area of interest.
The \emph{Task} (or task relevance~\cite{Lavrenko-PhD}) specifies the level of 
usefulness of the information found in an entity for the actual task performed by the user
(for example, writing a paper or preparing a lecture).
The final component, \emph{Context}, includes everything that is not previously covered by 
\emph{Topic} and \emph{Task}, but which nevertheless affects the whole process of retrieval (such as search costs, or
the amount of novel information found, or anything else). 

Since the information seeking process may evolve in time, 
a fourth dimension, \emph{Time}, 
is needed to model the fact that users often change their perception of the information they seek to find.
For example, at a certain point in time an information unit (\emph{Surrogate} or \emph{Document}) 
may not be relevant to a user request (\emph{Query} or \emph{Request}), 
however due to the evolving nature of the seeking process the user may learn something that would 
permit them to understand the content of the unit, which, in turn, may make the same unit relevant to the request.

A definition of relevance can, therefore, be seen as a point in the above four-dimensional space.
Mizzaro~\cite{Mizzaro} argues that the above framework can be used to model and compare 
different definitions of relevance. For example, the following expression may be used to model 
the system-oriented definition of relevance described in Section 3.1: 
\emph{Topical relevance} of a \emph{Surrogate} to a \emph{Query} at a certain point in \emph{Time} 
(the time when the request was formulated as a query and submitted to the retrieval system). 
However, finding an expression that may be used to model the INEX definition of relevance 
turns out to be quite a challenging task.
The main problem is that
both the INEX relevance dimensions, \emph{Exhaustivity} and \emph{Specificity},
are based on topical relevance, 
which corresponds to the \emph{Topic} component of the third relevance dimension in the unified framework.
We contend that one relevance dimension based on topical relevance should be used, 
or possibly two orthogonal dimensions that correspond to different components of the above framework.
In Section 6 we propose a much simpler definition of relevance, 
and argue that its corresponding relevance scale would be a better choice for INEX.

\section{Behaviour Analysis for \\Background Topics}

In this section, we separately analyse the assessor's and users' behaviour when judging the relevance of returned elements
for the \emph{Background} topic B1. In order to identify the best retrieval elements for this topic,
we also analyse and compare the level of agreement between the assessor and the users.

\subsection{Analysis of Assessor's Behaviour}

\begin{figure}[tb]
\epsfxsize=8.5cm
\centering\epsfbox{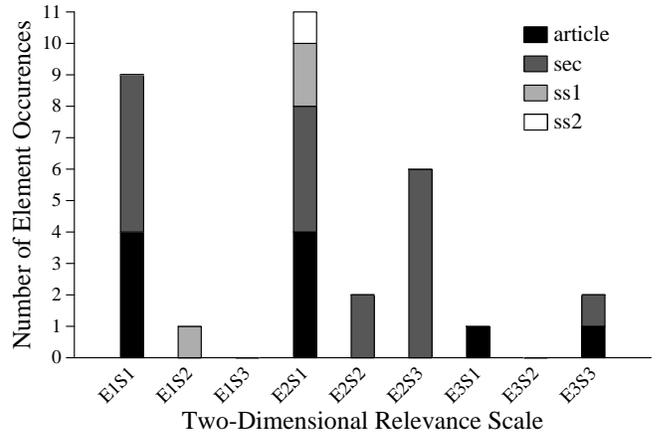}
\caption{Analysis of assessor's behaviour for topic B1. For each point of the relevance scale, the figure shows the total 
number of relevant elements, and the number of relevant elements for each of the element names.}
\label{B1-assessor}
\end{figure}

Figure~\ref{B1-assessor} shows an analysis of the 
relevance judgements for topic B1 (the INEX 2004 CO topic 192) that were obtained from one assessor.
As shown in the figure, we use only four element names in our analysis: \verb+article+, \verb+sec+, \verb+ss1+, and \verb+ss2+.
The $x$-axis contains the 9-point relevance scale which is a result of combining the grades of the two INEX relevance dimensions
(the case \verb+E0S0+ is not shown). 
The $y$-axis contains the number of occurrences of relevant elements for each point of the relevance scale.
For a relevance point, the number of occurrences of each of the four element names is also shown. 

The total number of relevant elements for topic B1 is 32. 
Of these, 11 elements have been judged as \verb+E2S1+, nine as \verb+E1S1+, six as \verb+E2S3+, 
two as \verb+E3S3+ or \verb+E2S2+, and one as \verb+E3S1+ or \verb+E1S2+.
Interestingly, none of the relevant elements have been judged as either \verb+E3S2+ or \verb+E1S3+.
The number of occurrences of the four element names is as follows.
The \verb+sec+ elements occur most frequently with~18 occurrences,
followed by \verb+article+ with ten, \verb+ss1+ with three, and \verb+ss2+ with one occurrence, respectively.
The total number of elements that have been judged as non-relevant (\verb+E0S0+) for topic B1 is 1158,
of which 513 are \verb+sec+ elements, 411 are \verb+ss1+, 186 are \verb+article+, and 48 are \verb+ss2+ elements.

\subsubsection*{Level of Overlap}
The above statistics show that the \verb+E2S1+ and \verb+E1S1+ 
points of the relevance scale contain around 63\% of the relevant elements for topic B1. 
Moreover, further analysis reveals that there is a substantial amount of overlap 
among these elements. 
More precisely, there is 64\% set-based overlap among the 11 \verb+E2S1+ elements,
where the four \verb+article+ elements contain all of the section and sub-section elements.
Similarly, there is 56\% overlap among \verb+E1S1+ elements,
where of nine elements, four \verb+article+ elements contain the other five \verb+sec+ elements.
Interestingly, the other points of the relevance scale do not suffer from overlap.
The two highly relevant elements (\verb+E3S3+), for example, belong to different XML files.

\subsubsection*{Correlation between Relevance Grades}
In the following we investigate the correlation between the grades of the two relevance dimensions for topic B1.
We want to check whether, while judging relevant elements, the assessor's choice of combining the grades
of the two relevance dimensions is influenced by a common aspect~\cite{Kazai-ECIR04}.

\begin{table*}[tp]
\begin{small}
\begin{center}
\begin{tabular}{c c c c c c c c c c c c}
\hline \hline
 & & & & \multicolumn{8}{c}{\bf{Exhaustivity}}  \\
\cline{5-12}
 & & & & \multicolumn{2}{c}{\bf{E3}} & & \multicolumn{2}{c}{\bf{E2}} & & \multicolumn{2}{c}{\bf{E1}} \\
\cline{5-6} \cline{8-9} \cline{11-12}
\bf{Assessor:} & & \multicolumn{1}{c}{\bf{Specificity}} & & \verb+Sp|Ex+ (\%) & \verb+Ex|Sp+ (\%)& & \verb+Sp|Ex+ (\%) & \verb+Ex|Sp+ (\%) & & \verb+Sp|Ex+ (\%) & \verb+Ex|Sp+ (\%) \\
\cline{3-12}
 & & \emph{S3} & & \bf{66.67} & 25.00 & & 31.57 & \emph{75.00} & & 0.00 & 0.00 \\ 
 & & \emph{S2} & & 0.00 & 0.00 & & 10.53 & \emph{66.67} & & 10.00 & 33.33 \\ 
 & & \emph{S1} & & 33.33 & 4.62 & & \bf{57.90} & \emph{52.38} & & \bf{90.00} & 43.00 \\ 
\hline \hline
& & & & \multicolumn{8}{c}{\bf{Exhaustivity}}  \\
\cline{5-12}
 & & & & \multicolumn{2}{c}{\bf{E3}} & & \multicolumn{2}{c}{\bf{E2}} & & \multicolumn{2}{c}{\bf{E1}} \\
\cline{5-6} \cline{8-9} \cline{11-12}
\bf{Users:} & & \multicolumn{1}{c}{\bf{Specificity}} & & \verb+Sp|Ex+ (\%) & \verb+Ex|Sp+ (\%)& & \verb+Sp|Ex+ (\%) & \verb+Ex|Sp+ (\%) & & \verb+Sp|Ex+ (\%) & \verb+Ex|Sp+ (\%) \\
\cline{3-12}
 & & \emph{S3} & & \bf{69.62} & \emph{69.62} & & 36.84 & 22.15 & & 12.26 & 8.23 \\ 
 & & \emph{S2} & & 27.22 & \emph{41.34} & & \bf{40.00} & 36.54 & & 21.70 & 22.12 \\ 
 & & \emph{S1} & & 3.16 & 5.16 & & 23.16 & 22.68 & & \bf{66.04} & \emph{72.16} \\ 
\hline \hline
\end{tabular}
\end{center}
\end{small}
\caption{Correlation between the grades of the two relevance dimensions for topic B1, as judged by both the assessor and the users.
Depending on the relevance dimension, the highest correlation of each grade is shown either in bold (for Exhaustivity) or italics (for Specificity).}
\label{B1-correlation}
\end{table*}

The top half of Table~\ref{B1-correlation} shows the correlation between the grades of the two relevance dimensions for topic B1, 
as judged by the assessor. 
For each grade of the \emph{Exhaustivity} relevance dimension (columns),
the value of \verb+Sp|Ex+ shows the percentage of the cases where an element is judged as \verb+Sp+ (specific),
given that it has already been judged as \verb+Ex+ (exhaustive). Similarly, for each grade of the \emph{Specificity} relevance dimension (rows),
the value of \verb+Ex|Sp+ shows the percentage of the cases where an element is judged as \verb+Ex+ (exhaustive),
given that it has already been judged as \verb+Sp+ (specific).
For example, the \verb+Sp|Ex+ value of column \verb+E3+ and row \verb+S3+ is 66.67, indicating that 
in 66.67\% of the cases a highly exhaustive element is also judged as highly specific.
We now analyse the correlation between the grades of each separate relevance dimension.

\begin{figure}[tb]
\epsfxsize=8.5cm
\centering\epsfbox{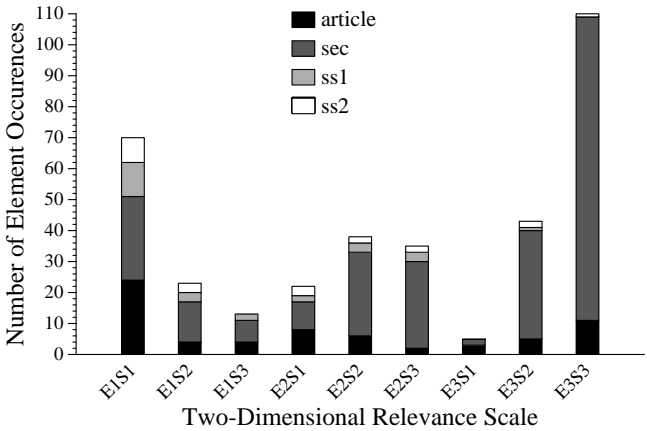}
\caption{Analysis of users' behaviour for topic B1. For each point of the relevance scale, the figure shows the total 
number of relevant elements, and the number of relevant elements for each of the element names.}
\label{B1-users}
\end{figure}

For \emph{Exhaustivity}, we observe that in 90\% of the cases 
a marginally exhaustive (\verb+E1+) element is also judged as margi-nally specific (\emph{S1}).
This is somehow intuitive, since 
by definition 
a marginally exhaustive element discusses only a few aspects of the information need,
so its focus may be on aspects other than those concerning the information need.
However, for topic B1, the number of \verb+E1+ elements is around 30\% of the total number of relevant elements, 
so the above correlation should be treated carefully.
In contrast, the number of fairly exhaustive elements (\verb+E2+) is around 60\% of the total number of relevant elements,
and in 58\% of the cases a fairly exhaustive element is (again) judged as \emph{S1}.
For highly exhaustive (\verb+E3+) elements, we find that in 67\% of the cases 
an \verb+E3+ element is also judged as highly specific (\emph{S3}), 
although the number of \verb+E3+ elements is very low (only 10\% of the total number of relevant elements).

For \emph{Specificity}, 
the number of marginally specific (\emph{S1}) elements is around 66\% of the total number of relevant elements,
where in 52\% of the cases an \emph{S1} element is judged as fairly exhaustive (\verb+E2+), while
in 43\% of the cases it is judged as marginally exhaustive (\verb+E1+).
Fairly specific (\emph{S2}) elements are 9\% of the total number of relevant elements,
and in 67\% of the cases an \emph{S2} element is judged as \verb+E2+.
Finally, in 75\% of the cases a highly specific (\emph{S3})
element is (again) judged as \verb+E2+, 
although the number of highly specific elements is around 25\% of the total number of relevant elements.

\subsection{Analysis of Users' Behaviour}

Figure~\ref{B1-users} shows the 
relevance judgements for topic B1 that were obtained from 50 users.
Unlike in the assessor's case, an element may have been judged by more than one user,
so each relevance point in Figure~\ref{B1-users} may contain multiple occurrences of a given element.

The total number of occurrences of relevant elements for topic B1 is 359. 
Around 61\% of this number are elements that have been judged either as \verb+E3S3+ (110),
\verb+E1S1+ (70), or \verb+E2S2+ (38).
All the 10 points of the relevance scale were used by users.
However, different number of users have judged elements for each relevance point.
For example, 41 (out of 50) users have judged at least one element as \verb+E3S3+, whereas 
this number is 35 for \verb+E1S1+, 23 for \verb+E2S2+, and 20 and below 
for the other points of the relevance scale. 
The \verb+sec+ elements occur most frequently with~246 occurrences, 
followed by \verb+article+ with 67, 
\verb+ss1+ with 25, 
and \verb+ss2+ with 21 
occurrences, respectively.
The total number of element occurrences judged as non-relevant (\verb+E0S0+) for topic B1 is 181, 
of which 80 are \verb+sec+ elements, 72 are \verb+article+, 26 are \verb+ss1+, and only 3 are \verb+ss2+ elements.
Also, 39 (out of 50) users have judged at least one element as \verb+E0S0+.

\begin{table*}[tp]
\begin{small}
\begin{center}
\begin{tabular}{c c c c c c c c c c c c c r r}
\hline
 \multicolumn{2}{c}{\bf{Assessor}} & & \multicolumn{11}{c}{\bf{User judgements}} & {\bf Agreement} \\
\cline{1-2} \cline{4-14} \cline{15-15}
Judgement & Total & & E3S3 & E3S2 & E3S1 & E2S3 & E2S2 & E2S1 & E1S3 & E1S2 & E1S1 & E0S0 & Total & (\%) \\
\hline \hline
\verb+E3S3+ & 2 & & 25 & 10 & 0 & 5 & 4 & 1 & 0 & 2 & 1 & 0 & 48 (2) & 52.08 \\ 
\verb+E3S2+  & 0 & & 0 & 0 & 0 & 0 & 0 & 0 & 0 & 0 & 0 & 0 & 0 (0) & 0.00 \\ 
\verb+E3S1+ & 1 & & 1 & 0 & 0 & 0 & 0 & 2 & 0 & 0 & 0 & 0 & 3 (1) & 0.00 \\ 
\verb+E2S3+ & 6 & & 60 & 14 & 1 & 18 & 13 & 4 & 3 & 7 & 8 & 0 & 128 (6) & 14.06 \\ 
\verb+E2S2+ & 2 & & 14 & 4 & 1 & 2 & 1 & 0 & 1 & 0 & 0 & 1 & 24 (1) & 4.17 \\ 
\verb+E2S1+ & 11 & & 1 & 0 & 0 & 2 & 1 & 0 & 0 & 0 & 2 & 0 & 6 (3) & 0.00 \\ 
\verb+E1S3+ & 0 & & 0 & 0 & 0 & 0 & 0 & 0 & 0 & 0 & 0 & 0 & 0 (0) & 0.00 \\ 
\verb+E1S2+ & 1 & & 0 & 0 & 0 & 2 & 1 & 0 & 0 & 0 & 1 & 0 & 4 (1) & 0.00 \\ 
\verb+E1S1+ & 9 & & 3 & 2 & 1 & 1 & 7 & 3 & 0 & 2 & 11 & 17 & 47 (5) & 23.40 \\ 
\verb+E0S0+ & 1158 & & 1 & 6 & 2 & 2 & 7 & 9 & 7 & 6 & 36 & 99 & 175 (59) & 56.57 \\ 
\hline
\verb+Total+ & 1190 & & 105 & 36 & 5 & 32 & 34 & 19 & 11 & 17 & 59 & 117 & 435 (78) & \bf{15.10} \\ 
\hline \hline
\end{tabular}
\end{center}
\end{small}
\caption{The level of agreement between the assessor and the users for topic B1. For each point of the relevance scale, the percentage of users that agree with the assessor's judgements of corresponding elements is shown. Numbers in brackets represent numbers of unique elements judged by users. The overall level of agreement for topic B1 is shown in bold.}
\label{B1-10-point-agreement}
\end{table*}

\subsubsection*{Level of Overlap}
A more detailed analysis of the user judgements for topic B1 reveals that there is almost no overlap 
among the elements that belong to any of the nine points of the relevance scale. 
More precisely, there is 14\% set-based overlap among the 110 \verb+E3S3+ elements,
0\% overlap among the 70 \verb+E1S1+ elements,
and 0\% overlap for the other seven points of the relevance scale.
The above finding therefore confirms the hypothesis that users do not want to retrieve (and thus do not tolerate) redundant information. 

\subsubsection*{Correlation between Relevance Grades}
The lower half of Table~\ref{B1-correlation} shows the correlation between the grades of the two relevance dimensions for topic B1, 
as judged by users.
For both \emph{Exhaustivity} and \emph{Specificity}, 
two strong correlations are visible.
First, 
in 66\% of the cases 
a marginally exhaustive (\verb+E1+) element 
is also judged as marginally specific (\emph{S1}) (and vice versa).
Second, 
in 70\% of the cases a highly exhaustive (\verb+E3+) element 
is also judged as highly specific (\emph{S3}) (and vice versa).
The number of \verb+E1+ elements is around 30\% of the total number of relevant elements, 
whereas 44\% of the total number of relevant elements are \verb+E3+ elements.
The number of \emph{S1} and \emph{S3} elements is almost the same as the number of 
\verb+E1+ and \verb+E3+ elements, respectively.
No strong correlations are, however, visible in the case of \verb+E2+ and \emph{S2} elements.

\subsection{Analysis of the Level of Agreement}

The analysis of the level of agreement concerns the amount of 
information identified as relevant by \emph{both} the assessor and the users.
The aim of this analysis is to identify the best units of retrieval for topic B1.

\begin{table*}[tp]
\begin{small}
\begin{center}
\begin{tabular}{l c c c c c c c c c c c c c c}
\hline \hline
\multicolumn{2}{c}{\bf{File: cg/1998/g1016}} & & & & & & & & & & & & & \\  
 & & & & & & & & & & & & & & \\  
 \multicolumn{2}{c}{\bf{Assessor}} & & \multicolumn{10}{c}{\bf{User judgements}} & & \bf{Total} \\
\cline{1-2} \cline{4-13} \cline{15-15}
\multicolumn{1}{c}{Element} & Judgement & & E3S3 & E3S2 & E3S1 & E2S3 & E2S2 & E2S1 & E1S3 & E1S2 & E1S1 & E0S0 & & (users) \\
\hline 
\verb+/article[1]+ & \verb+E3S3+ & & 9 & 3 & 0 & 0 & 0 & 0 & 0 & 0 & 0 & 0 & & 12 \\ 
\verb+//bdy[1]/sec[2]+ & \verb+E2S3+ & & 9 & 5 & 1 & 7 & 6 & 2 & 1 & 2 & 2 & 0 & & 35 \\ 
\verb+//bdy[1]/sec[3]+ & \verb+E2S2+ & & 14 & 4 & 1 & 2 & 1 & 0 & 1 & 0 & 0 & 1 & & 24 \\ 
\verb+//bdy[1]/sec[4]+ & \verb+E2S3+ & & 19 & 0 & 0 & 4 & 1 & 1 & 0 & 0 & 2 & 0 & & 27 \\ 
\verb+//bdy[1]/sec[5]+ & \verb+E2S3+ & & 18 & 3 & 0 & 3 & 2 & 1 & 0 & 2 & 1 & 0 & & 30 \\ 
\verb+//bdy[1]/sec[6]+ & \verb+E2S3+ & & 8 & 2 & 0 & 2 & 1 & 0 & 1 & 0 & 1 & 0 & & 15 \\ 
\verb+//bdy[1]/sec[7]+ & \verb+E2S3+ & & 6 & 4 & 0 & 2 & 2 & 0 & 1 & 3 & 2 & 0 & & 20 \\ 
\hline \hline
 & & & & & & & & & & & & & & \\  
\multicolumn{2}{c}{\bf{File: cg/1995/g5095}} & & & & & & & & & & & & & \\  
 & & & & & & & & & & & & & & \\  
 \multicolumn{2}{c}{\bf{Assessor}} & & \multicolumn{10}{c}{\bf{User judgements}} & & \bf{Total} \\
\cline{1-2} \cline{4-13} \cline{15-15}
\multicolumn{1}{c}{Element} & Judgement & & E3S3 & E3S2 & E3S1 & E2S3 & E2S2 & E2S1 & E1S3 & E1S2 & E1S1 & E0S0 & & (users) \\
\hline 
\verb+/article[1]+ & \verb+E3S1+ & & 1 & 0 & 0 & 0 & 0 & 2 & 0 & 0 & 0 & 0 & & 3 \\ 
\verb+//bdy[1]/sec[1]+ & \verb+E3S3+ & & 16 & 7 & 0 & 5 & 4 & 1 & 0 & 2 & 1 & 0 & & 36 \\ 
\verb+//bdy[1]/sec[2]+ & \verb+E0S0+ & & 0 & 0 & 0 & 0 & 0 & 0 & 0 & 0 & 1 & 2 & & 3 \\ 
\verb+//bdy[1]/sec[3]+ & \verb+E0S0+ & & 0 & 0 & 0 & 0 & 0 & 0 & 0 & 0 & 0 & 2 & & 2 \\ 
\verb+//bdy[1]/sec[4]+ & \verb+E0S0+ & & 0 & 0 & 0 & 0 & 0 & 0 & 0 & 0 & 0 & 3 & & 3 \\ 
\hline \hline
\end{tabular}
\end{center}
\end{small}
\caption{Distribution of relevance judgements for the XML files \emph{cg/1998/g1016} (top) and \emph{cg/1995/g5095} (bottom) for topic B1. For each element, the assessor judgement and the distribution of users' judgements are shown. The total number of users who judged a particular element is listed in the last column.}
\label{Relevant-per-file}
\end{table*}

Table~\ref{B1-10-point-agreement} shows the level of agreement between the assessor and the users for each point of the relevance scale. 
The two columns on the left refer to the assessor's judgements, where for each relevance point (the \verb+Judgement+ column),
the total number of judged elements that belong to this point is shown (the \verb+Total+ column). 
The values in the \verb+User Judgements+ columns show how users actually judged any (or all) of the corresponding elements judged by the assessor. The \verb+Total+ column on the right shows the total number of user judgements for each point of the relevance scale.
Numbers in brackets represent numbers of unique elements judged by users.
The \verb+Agreement+ column shows the level of agreement between the assessor and the users, where the percentage is calculated 
for each relevance point.

For example, the first row in the table indicates that there are two elements judged as \verb+E3S3+ by the assessor, and that of 48 total user judgements, 
there are 25 cases when users judged any (or both) of these two elements as \verb+E3S3+, ten cases as \verb+E3S2+, five cases as \verb+E2S3+, and so on. 
The level of agreement between the assessor and the users for the \verb+E3S3+ point of the relevance scale is
52.08\% (since in 25 out of 48 cases users judged these elements as \verb+E3S3+).
Note that for this relevance point we only consider the user judgements made on two unique elements, 
which correspond to the same elements judged as \verb+E3S3+ by the assessor.
As shown in the table, the overall level of agreement 
between the assessor and the users for topic B1 is 15\%.

Several observations can be made from the statistics shown in Table~\ref{B1-10-point-agreement}.

First, users judged 19 (unique) of the 32 \emph{relevant} elements as identified by the assessor for topic B1.
In 7\% of the cases, however, users judged some of these elements to be \emph{not relevant}.
Conversely, 59 (unique) of the 1,158 \emph{non-relevant} elements, as identified by the assessor,
were also judged by users, and in 43\% of the cases users judged some of those elements to be \emph{relevant}.

Second, the highest level of agreement between the assessor and the users 
is on highly relevant (\verb+E3S3+) and non-relevant (\verb+E0S0+) elements, 
with agreement values of 52\% and 57\%, respectively.
This shows that both the assessor and the users clearly perceive the end points of the relevance scale.
However, the other points of the relevance scale are not perceived as well. 
For example, although the highest number of user judgements is on the \verb+E2S3+ relevance point (around 50\%),
in only 14\% of the cases users actually judged these elements as \verb+E2S3+. In fact, in the 
majority of the cases (47\%), the users judged these elements to be highly relevant (\verb+E3S3+). 
Similar observations can be made for the \verb+E1S1+ relevance point, 
where in 36\% of the cases the users judged these elements to be non-relevant (\verb+E0S0+). 
Note that, even though the number of judged \verb+E3S3+ and \verb+E1S1+ elements is roughly the same, 
the level of agreement for the \verb+E3S3+ relevance point is more than two times greater than the level
of agreement for the \verb+E1S1+ relevance point.

Last, a more detailed analysis of the above statistics 
reveals that the agreement between the assessor and the users 
is almost the same for each separate relevance dimension. 
More precisely, the overall agreement 
for \emph{Exhaustivity} is 45\%, whereas the overall agreement for \emph{Specificity} is 44\%.
The agreement for highly exhaustive (\verb+E3+) elements is 71\%,
where 20\% of the total number of confirmed \emph{relevant} elements is on \verb+E3+ elements.
On the other hand, the agreement for highly specific (\verb+S3+) elements is 63\%,
where 68\% of the confirmed relevant elements are \verb+S3+ elements.
This shows that although the number of user judgements for the \verb+S3+ grade
is more than three times greater than the number of judgements for the \verb+E3+ grade,
highly exhaustive elements are perceived better than highly specific elements.

\subsubsection*{Best Units of Retrieval}

Previous analysis shows that of
all the \emph{relevant} elements as judged by users,
the \verb+E3S3+ point of the relevance scale has the highest level of agreement.
There are two elements judged as highly relevant by the assessor for topic B1 -- one \verb+article+ and one \verb+sec+ -- 
that belong to different XML files.
The \verb+article+ element belongs to file \verb+cg/1998/g1016+,
while the \verb+sec+ element belongs to \verb+cg/1995/g5095+.
We are interested in finding in these files the \emph{best units of retrieval} for topic B1.
In the following analysis, we examine the retrieval behaviour of both the assessor and the users for each of these files.

Table~\ref{Relevant-per-file} shows the distribution of relevance judgements
for relevant elements in the two XML files,
as done by both the assessor and the users. The two columns on the left refer to the assessor, 
where for each relevant element in the file (the \verb+Element+ column),
the assessor's judgement is also shown (the \verb+Judgement+ column).
The values in the \verb+User Judgements+ columns 
show the distribution of users' judgements for each particular element; that is,
the number below each relevance point represents the number of users that judged that element.
The total number of users who judged a particular element is shown in the \verb+Total+ column.

For the file \verb+cg/1998/g1016+, 
the top half of the table shows that 
the highly relevant (\verb+E3S3+) \verb+article+ element was judged by 12 (out of 50) 
users, and that 75\% of them confirmed it to also be highly relevant.
Interestingly, around 70\% of the relevant elements in this file have been judged as \verb+E2S3+ by the assessor, 
and there were 25 users (on average) who have also judged these elements. 
However, there is only a 14\% agreement (on average) between the assessor and the users for the \verb+E2S3+ relevance point.
In fact, if we take a closer look at the user judgements, we see that 
most users judged the \verb+E2S3+ elements to be highly relevant (\verb+E3S3+) elements.
For example, there were 27 users in total who judged the \verb+sec[4]+ element (judged as \verb+E2S3+ by the assessor), 
and 70\% of them judged this element to be highly relevant (\verb+E3S3+).

The above analysis shows that the agreement between the users and the assessor on the 
\emph{best units of retrieval} for the file \verb+cg/1998/g1016+ is not exact. 
Further analysis 
confirms that
the level of agreement between the assessor and the users is greater for highly exhaustive elements than for highly specific ones.
More precisely, although the number of user judgements for the \verb+S3+ grade
is more than ten times greater than the number of judgements for the \verb+E3+ grade,
there is a 65\% agreement for highly specific elements,
while there is a 100\% agreement for highly exhaustive elements.

For the file \verb+cg/1995/g5095+,
the lower half of Table~\ref{Relevant-per-file} shows that 
there are only two elements 
identified as \emph{relevant} by the assessor,
which makes it impossible to draw any sound conclusions. 
The highly relevant \verb+sec+ element was judged by 36 (out of 50) 
users, and around 45\% of the users also confirmed it to be highly relevant.
Interestingly, three \verb+sec+ elements were judged as not relevant by the assessor, 
and almost all of the users who judged these elements also confirm them to be non-relevant.

\section{Behaviour Analysis for \\Comparison Topics}

\begin{figure}[tb]
\epsfxsize=8.5cm
\centering\epsfbox{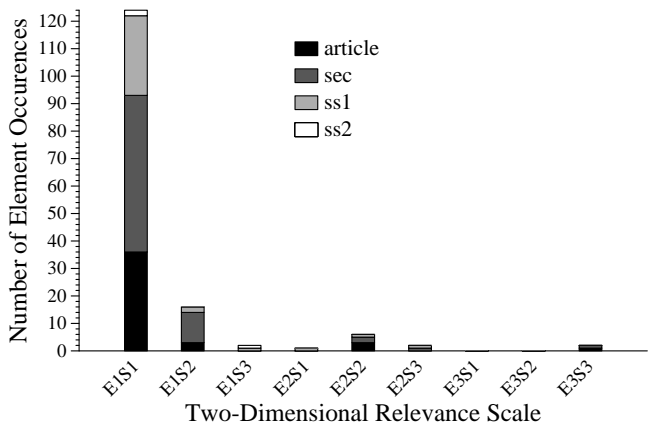}
\caption{Analysis of assessor's behaviour for topic C2. For each point of the relevance scale, the figure shows the total 
number of relevant elements, and the number of relevant elements for each of the element names.}
\label{C2-assessor}
\end{figure}

In this section, we separately analyse the assessor's and users' behaviour when judging the relevance of returned elements
for the \emph{Comparison} topic C2. 
In order to identify the best retrieval elements for this topic,
we also analyse and compare the level of agreement between the assessor and the users.

\subsection{Analysis of Assessor's Behaviour}

\begin{table*}[tp]
\begin{small}
\begin{center}
\begin{tabular}{c c c c c c c c c c c c}
\hline \hline
 & & & & \multicolumn{8}{c}{\bf{Exhaustivity}}  \\
\cline{5-12}
 & & & & \multicolumn{2}{c}{\bf{E3}} & & \multicolumn{2}{c}{\bf{E2}} & & \multicolumn{2}{c}{\bf{E1}} \\
\cline{5-6} \cline{8-9} \cline{11-12}
\bf{Assessor:} & & \multicolumn{1}{c}{\bf{Specificity}} & & \verb+Sp|Ex+ (\%) & \verb+Ex|Sp+ (\%)& & \verb+Sp|Ex+ (\%) & \verb+Ex|Sp+ (\%) & & \verb+Sp|Ex+ (\%) & \verb+Ex|Sp+ (\%) \\
\cline{3-12}
 & & \emph{S3} & & \bf{100.00} & 33.33 & & 22.22 & 33.33 & & 1.41 & 33.33 \\ 
 & & \emph{S2} & & 0.00 & 0.00 & & \bf{66.67} & 27.27 & & 11.27 & \emph{72.73} \\ 
 & & \emph{S1} & & 0.00 & 0.00 & & 11.11 & 0.80 & & \bf{87.32} & \emph{99.20} \\ 
\hline \hline
& & & & \multicolumn{8}{c}{\bf{Exhaustivity}}  \\
\cline{5-12}
 & & & & \multicolumn{2}{c}{\bf{E3}} & & \multicolumn{2}{c}{\bf{E2}} & & \multicolumn{2}{c}{\bf{E1}} \\
\cline{5-6} \cline{8-9} \cline{11-12}
\bf{Users:} & & \multicolumn{1}{c}{\bf{Specificity}} & & \verb+Sp|Ex+ (\%) & \verb+Ex|Sp+ (\%)& & \verb+Sp|Ex+ (\%) & \verb+Ex|Sp+ (\%) & & \verb+Sp|Ex+ (\%) & \verb+Ex|Sp+ (\%) \\
\cline{3-12}
 & & \emph{S3} & & \bf{52.99} & \emph{48.84} & & 30.13 & 32.56 & & 13.77 & 18.60 \\ 
 & & \emph{S2} & & 35.04 & 27.33 & & \bf{43.59} & \emph{43.33} & & 27.54 & 29.33 \\ 
 & & \emph{S1} & & 11.97 & 8.50 & & 26.28 & 27.45 & & \bf{58.68} & \emph{64.05} \\ 
\hline \hline
\end{tabular}
\end{center}
\end{small}
\caption{Correlation between the grades of the two relevance dimensions for topic C2, as judged by both the assessor and the users.
Depending on the relevance dimension, the highest correlation of each grade is shown either in bold (for Exhaustivity) or italics (for Specificity).}
\label{C2-correlation}
\end{table*}

Figure~\ref{C2-assessor} shows the relevance judgements for the INEX 2004 CO topic 198 (topic C2) that were obtained from one assessor.
As shown in the figure, the total number of relevant elements for topic C2 is 153, of which
the majority (81\%) have been judged as \verb+E1S1+.
Interestingly, none of the relevant elements have been judged as either \verb+E3S2+ or \verb+E3S1+.
The distribution of the four element names is as follows.
The \verb+sec+ elements occur most frequently with~72 occurrences,
followed by \verb+article+ with 43, \verb+ss1+ with 35, and \verb+ss2+ with only three occurrences, respectively.
The total number of elements that have been judged as non-relevant (\verb+E0S0+) for topic C2 is 1094,
of which 547 are \verb+sec+ elements, 304 are \verb+ss1+, 191 are \verb+article+, and 52 are \verb+ss2+ elements.

\subsubsection*{Level of Overlap}
The above statistics show that the \verb+E1S1+ 
point of the relevance scale contains almost all of the relevant elements for topic C2. 
However, as for the topic B1, there is a substantial overlap 
among these elements. 
More precisely, there is a 63\% set-based overlap among the 124 \verb+E1S1+ elements.
On the other hand, the other points of the relevance scale -- except the \verb+E3S3+ point -- do not suffer from overlap.
For the \verb+E3S3+ point, there is a 50\% set-based overlap, where the two highly relevant elements (one \verb+article+ and one \verb+sec+) belong to the same XML file.

\newpage

\subsubsection*{Correlation between Relevance Grades}
The top half of Table~\ref{C2-correlation} shows the correlation between the grades of the two relevance dimensions for topic C2, 
as judged by the assessor. 
We observe that each of the three grades of the \emph{Exhaustivity} dimension
is strongly correlated with its corresponding grade of the \emph{Specificity} dimension.
This is most evident for the \verb+E1+ grade, where 
in 87\% of the cases a marginally exhaustive (\verb+E1+) element is also judged as marginally specific (\emph{S1}).
The number of \verb+E1+ elements is 93\% of the total number of relevant elements.
The same is not true for the grades of the Specificity dimension, however,
where both the \verb+S2+ and \verb+S1+ grades are strongly correlated with the \verb+E1+ grade.
Most notably, 
in 99\% of the cases a marginally specific (\verb+S1+) element is also judged as marginally exhaustive (\emph{E1}),
where the number of \verb+S1+ elements is 82\% of the total number of relevant elements.

\subsection{Analysis of Users' Behaviour}

\begin{figure}[tb]
\epsfxsize=8.5cm
\centering\epsfbox{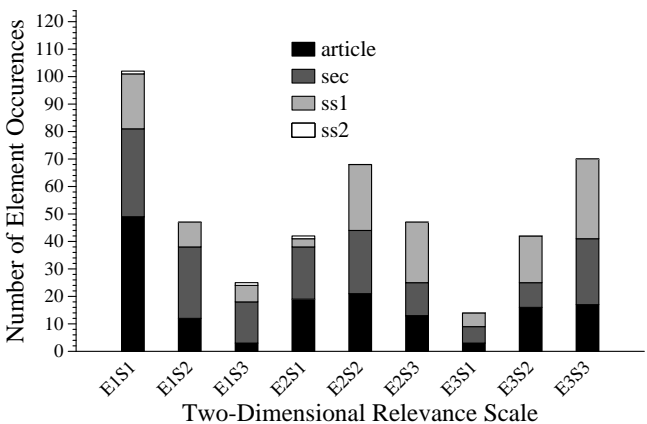}
\caption{Analysis of users' behaviour for topic C2. For each point of the relevance scale, the figure shows the total 
number of relevant elements, and the number of relevant elements for each of the element names.}
\label{C2-users}
\end{figure}

Figure~\ref{C2-users} shows the 
relevance judgements for topic C2 that were obtained from 52 users.
As shown in the figure, the total number of occurrences of relevant elements is 445, 
of which around half of that number are elements that belong to the following three points of the relevance scale:
\verb+E1S1+~(101), \verb+E2S2+ (66), and \verb+E3S3+ (63).
Interestingly, approximately the same number of users (34 out of 52) 
judged at least one element that belongs to each of these three points.
In contrast, 22 users (on average) judged at least one element that belongs to the other six points of the relevance scale.

The distribution of the four element names is as follows.
The \verb+sec+ and \verb+article+ elements occur most frequently with~159 and 153 occurrences, 
followed by \verb+ss1+ elements with 130, 
and \verb+ss2+ elements with only three occurrences, respectively. 
The total number of element occurrences judged as non-relevant (\verb+E0S0+) for topic C2 is 170, 
of which 116 are \verb+sec+ elements, 27 are \verb+ss1+, 26 are \verb+article+, and only one element is an \verb+ss2+ element.
Also, 38 out of 52 users have judged at least one element as \verb+E0S0+.

\begin{table*}[tp]
\begin{small}
\begin{center}
\begin{tabular}{c c c c c c c c c c c c c r r}
\hline
 \multicolumn{2}{c}{\bf{Assessor}} & & \multicolumn{11}{c}{\bf{Users}} & \bf{Agreement} \\
\cline{1-2} \cline{4-14} \cline{15-15}
Relevance & Total & & E3S3 & E3S2 & E3S1 & E2S3 & E2S2 & E2S1 & E1S3 & E1S2 & E1S1 & E0S0 & Total & (\%) \\
\hline \hline
\verb+E3S3+ & 2 & & 6 & 4 & 1 & 1 & 0 & 0 & 1 & 0 & 1 & 0 & 14 (2) & 42.86 \\ 
\verb+E3S2+ & 0 & & 0 & 0 & 0 & 0 & 0 & 0 & 0 & 0 & 0 & 0 & 0 (0) & 0.00 \\ 
\verb+E3S1+ & 0 & & 0 & 0 & 0 & 0 & 0 & 0 & 0 & 0 & 0 & 0 & 0 (0) & 0.00 \\ 
\verb+E2S3+ & 2 & & 16 & 4 & 1 & 6 & 7 & 1 & 0 & 0 & 2 & 1 & 38 (2) & 15.79 \\ 
\verb+E2S2+ & 6 & & 20 & 8 & 0 & 12 & 12 & 6 & 3 & 5 & 7 & 3 & 76 (5) & 15.79 \\ 
\verb+E2S1+ & 1 & & 2 & 0 & 1 & 1 & 4 & 0 & 0 & 1 & 1 & 0 & 10 (1) & 0.00 \\ 
\verb+E1S3+ & 2 & & 0 & 0 & 0 & 0 & 1 & 0 & 1 & 0 & 1 & 1 & 4 (2) & 25.00 \\ 
\verb+E1S2+ & 16 & & 1 & 1 & 1 & 2 & 1 & 0 & 1 & 2 & 1 & 2 & 12 (7) & 16.67 \\ 
\verb+E1S1+ & 124 & & 17 & 19 & 6 & 16 & 24 & 16 & 8 & 18 & 45 & 38 & 207 (34) & 21.74 \\ 
\verb+E0S0+ & 1094 & & 2 & 2 & 2 & 3 & 3 & 10 & 5 & 9 & 25 & 85 & 146 (52) & 58.22 \\ 
\hline
\verb+Total+ & 1247 & & 64 & 38 & 12 & 41 & 52 & 33 & 19 & 35 & 83 & 130 & 507 (105) & \bf{19.61} \\ 
\hline \hline
\end{tabular}
\end{center}
\end{small}
\caption{The level of agreement between the assessor and the users for topic C2. For each point of the relevance scale, the percentage of users that agree with the assessor's judgements of corresponding elements is shown. Numbers in brackets represent numbers of unique elements judged by users. The overall level of agreement for topic C2 is shown in bold.}
\label{C2-10-point-agreement}
\end{table*}

\subsubsection*{Level of Overlap}
Further analysis of the user judgements for topic C2 reveals that there is almost no overlap 
among the elements that belong to any of the nine points of the relevance scale. 
More specifically, there is 3\% set-based overlap for the \verb+E1S1+ point,
0\% for the \verb+E2S2+ point, 9\% overlap for the \verb+E3S3+ point,
and 0\% overlap for the other six points of the relevance scale.

\subsubsection*{Correlation between Relevance Grades}
The lower half of Table~\ref{C2-correlation} shows the correlation between the grades of the two relevance dimensions for topic C2, 
as judged by users.
Although no strong correlations are visible, 
the values in the table show that, as in assessor's case,
the highest correlations are between the same grades of each of the two relevance dimensions.

\subsection{Analysis of the Level of Agreement}

In this section we analyse the amount of 
information identified as relevant by \emph{both} the assessor and the users.
Table~\ref{C2-10-point-agreement} shows the level of agreement between the assessor and the users for each point of the relevance scale. 
Three observations can be made from the statistics shown in the table.

First, users judged 53 (unique) of the 153 \emph{relevant} elements as identified by the assessor for topic C2.
In 12\% of the cases, however, users judged these elements to be \emph{not relevant}.
Conversely, 52 (unique) of the 1094 \emph{non-relevant} elements, as identified by the assessor,
were also judged by users, and in 42\% of the cases users judged these elements to be \emph{relevant}.

Second, as for topic B1 the highest level of agreement between the assessor and the users 
is on the end points of the relevance scale: \verb+E3S3+ (43\%) and \verb+E0S0+ (58\%),
although the number of user judgements for the \verb+E3S3+ relevance point is much less than 
the number of judgements for the \verb+E0S0+ point.
The \verb+E1S1+ relevance point has the highest number of user judgements (207 out of 507),
and in 22\% of the cases users also judged these elements to be \verb+E1S1+. 
Also, there are 76 user judgements for the \verb+E2S2+ relevance point,
however in 26\% of the cases users actually judged the \verb+E2S2+ elements to be highly relevant (\verb+E3S3+) elements.

\begin{table*}[tp]
\begin{small}
\begin{center}
\begin{tabular}{l c c c c c c c c c c c c c c}
\hline \hline
\multicolumn{2}{c}{\bf{File: co/2000/rx023}} & & & & & & & & & & & & & \\  
 & & & & & & & & & & & & & & \\  
 \multicolumn{2}{c}{\bf{Assessor}} & & \multicolumn{10}{c}{\bf{User judgements}} & & \bf{Total} \\
\cline{1-2} \cline{4-13} \cline{15-15}
\multicolumn{1}{c}{Element} & Judgement & & E3S3 & E3S2 & E3S1 & E2S3 & E2S2 & E2S1 & E1S3 & E1S2 & E1S1 & E0S0 & & (users) \\
\hline 
\small\verb+/article[1]+ & \small\verb+E3S3+ & & 4 & 3 & 0 & 0 & 0 & 0 & 0 & 0 & 0 & 0 & & 7 \\ 
\small\verb+//bdy[1]/sec[1]+ & \small\verb+E2S2+ & & 5 & 0 & 0 & 1 & 2 & 2 & 1 & 3 & 1 & 2 & & 17 \\ 
\small\verb+//bdy[1]/sec[2]+ & \small\verb+E2S2+ & & 0 & 0 & 0 & 1 & 1 & 1 & 0 & 1 & 0 & 1 & & 5 \\ 
\small\verb+//bdy[1]/sec[3]+ & \small\verb+E3S3+ & & 2 & 1 & 1 & 1 & 0 & 0 & 1 & 0 & 1 & 0 & & 7 \\ 
\small\verb+//bdy[1]/sec[3]/ss1[1]+ & \small\verb+E2S2+ & & 10 & 3 & 0 & 8 & 6 & 2 & 2 & 1 & 1 & 0 & & 33 \\ 
\small\verb+//bdy[1]/sec[3]/ss1[2]+ & \small\verb+E2S1+ & & 2 & 0 & 1 & 1 & 4 & 0 & 0 & 1 & 1 & 0 & & 10 \\ 
\small\verb+//bdy[1]/sec[3]/ss1[4]+ & \small\verb+E1S1+ & & 3 & 0 & 1 & 1 & 1 & 0 & 0 & 0 & 1 & 0 & & 7 \\ 
\small\verb+//bdy[1]/sec[3]/ss1[5]+ & \small\verb+E2S3+ & & 7 & 3 & 1 & 4 & 2 & 0 & 0 & 0 & 0 & 0 & & 17 \\ 
\small\verb+//bdy[1]/sec[3]/ss1[6]+ & \small\verb+E1S3+ & & 0 & 0 & 0 & 0 & 1 & 0 & 0 & 0 & 0 & 1 & & 2 \\ 
\small\verb+//bdy[1]/sec[4]+ & \small\verb+E2S3+ & & 9 & 1 & 0 & 2 & 5 & 1 & 0 & 0 & 2 & 1 & & 21 \\ 
\small\verb+//bm[1]/app[1]/sec[1]+ & \small\verb+E1S1+ & & 0 & 0 & 0 & 1 & 0 & 0 & 0 & 1 & 0 & 3 & & 5 \\ 
\hline \hline
\end{tabular}
\end{center}
\end{small}
\caption{Distribution of relevance judgements for the XML file \emph{co/2000/rx023} for topic C2. For each element, the assessor judgement and the distribution of users' judgements are shown. The total number of users who judged a particular element is listed in the last column.}
\label{C2-Relevant-per-file}
\end{table*}

Third, a more detailed analysis shows that the level of agreement between the assessor and the users 
differs for each separate relevance dimension.
More precisely, the overall agreement 
for \emph{Exhaustivity} is 53\%, while the overall agreement for \emph{Specificity} is 45\%.
The agreement for highly exhaustive (\verb+E3+) elements is 79\%,
and 4\% of the total number of confirmed \emph{relevant} elements is on \verb+E3+ elements.
In contrast, the agreement for highly specific (\verb+S3+) elements is 55\%,
where 18\% of confirmed relevant elements are \verb+S3+ elements.
This shows that, as for topic B1, highly exhaustive elements are perceived better than highly specific 
elements.

\subsubsection*{Best Units of Retrieval}

There are two elements judged as highly relevant by the assessor for topic C2, one \verb+article+ and one \verb+sec+, 
which belong to the same XML file: \verb+co/2000/rx023+.
To identify the best units of retrieval, in the following we examine the behaviour of both the assessor and the users for this file.

Table~\ref{C2-Relevant-per-file} shows the distribution of relevance judgements
for relevant elements in the XML file \verb+co/2000/rx023+,
as done by both the assessor and the users. 
As shown in the table, the two highly relevant (\verb+E3S3+) elements were judged by the same number of users 
(seven out of 52). 
Of the users that judged each of these elements, 57\% confirmed the \verb+article[1]+ to be highly relevant, while
only 29\% confirmed the \verb+sec[3]+ element to be highly relevant.
Many users, however, found the child elements of the \verb+sec[3]+ element (such as \verb+ss1[1]+, \verb+ss1[4]+ and \verb+ss1[5]+) 
to be highly relevant.

From the above distribution of relevance judgements it is hard to draw any sound conclusions  
as to which elements constitute \emph{best units of retrieval} for this file. 
Further analysis of the two behaviours for this file
again confirms that, for topic C2, 
the level of agreement between the assessor and the users is greater for highly exhaustive than for highly specific elements.
Specifically, although 
the number of user judgements for the \verb+S3+ grade
is four times greater than the number of judgements for the \verb+E3+ grade,
the agreement for highly specific elements is 56\%,
while there is a 79\% agreement for highly exhaustive elements.

\section{Discussion}

In previous sections we 
separately studied the behaviour of the assessor and the users when judging the relevance of returned elements.
We also analysed the level of agreement between the assessor and the users
in order to identify the best units of retrieval for each of the two topics.

According to the assessor,
most of the relevant elements for topic B1 
reside in the \verb+E2S1+ and \verb+E1S1+ points of the relevance scale.
The \verb+E1S1+ relevance point also contains most of the relevant elements for topic C2.
In both topic cases, however,
there is a substantial overlap among these relevant elements:
60\% for topic B1, and 63\% for topic C2.
There are no visible correlations between the grades of each relevance dimension for the assessor of topic B1,
whereas for for the assessor of topic C2
each of the three grades of the \emph{Exhaustivity} dimension
is strongly correlated with its corresponding grade of the \emph{Specificity} dimension.

According to users, 
most of the relevant elements in both topic cases
reside in the \verb+E1S1+, \verb+E2S2+, and \verb+E3S3+ relevance points. 
Moreover, there is almost no overlap among the relevant elements.
Unlike in the assessor's case, 
the highest correlations between the grades of the relevance dimensions
are between the same grades of each of the two dimensions, irrespective of the choice of the topic used.
This shows that the two INEX relevance dimensions are not perceived as orthogonal dimensions; in fact, 
users behave as if each of the grades from either dimension belongs to only one relevance dimension. 

The latter finding suggests that the \emph{common aspect} influencing the choice of combining grades from the two INEX relevance dimensions 
is the fact that the users can not make a clear distinction between the two dimensions (since they are both based on topical relevance).
However, it does not mean that the two INEX relevance dimensions are the same. On the contrary, 
from the \emph{Exhaustivity} definition, higher aspect coverage does not imply that there is less non-relevant information in an element, 
which means there is no one-to-one correspondence between the two INEX dimensions.
Rather, the users' perception -- which was empirically identified in this study --
suggests that the cognitive load of simultaneously choosing the grades for \emph{Exhaustivity} and \emph{Specificity}
is too difficult a task.
Part of the problem may be that the users (and the assessor) 
may not have understood an important property of the \emph{Specificity} dimension:
an element should be judged as \emph{highly specific (S3)} if it \emph{does not} contain \emph{non-relevant} information.

The low level of overlap between the judged elements in the users' case shows that 
\emph{retrieving overlapping units of information is not what users really want}.
However, the higher level of overlap in the assessor's case does not necessarily mean that 
the assessor's behaviour is very different from that of users; indeed, there are \emph{at least} two external factors
that may have influenced the observed level of overlap for the assessor:

\begin{itemize}
\item The assessor was required to judge many more elements than the users, in order for the obtained relevance judgements 
to be as exhaustive (and as consistent) as possible; and
\item The assessor and the users used different system interfaces, which may have introduced a bias in the way the elements were judged.
\end{itemize}

The highest level of agreement between the assessor and the users in both topic cases 
is respectively on highly relevant (\verb+E3S3+) and non-relevant (\verb+E0S0+) elements, 
which shows that both the assessor and the users clearly perceive the end points of the relevance scale.
However, \emph{the other points of 10-point relevance scale were not perceived as well}.
When the two relevance dimensions were analysed separately,
we observed that -- in both topic cases -- \emph{Exhaustivity} is perceived better than \emph{Specificity}. 

The above findings suggest that a much simpler relevance scale, and therefore, a much simpler relevance definition,
would be a preferable choice for INEX.
In the following we propose one such definition of relevance. 

\subsection*{Aspects and Dimensions of Relevance}

There are three aspects on which our new definition of relevance is based on:

\begin{itemize}
\item There should be only \emph{one} dimension of relevance based on \emph{topical relevance} (rather than two);
\item The relevance dimension should use a \emph{binary} relevance scale (rather than graded relevance scale), which determines
whether a unit of information is \emph{relevant} or \emph{not} to an information need; and
\item There should be second \emph{orthogonal} dimension of relevance, based on the hierarchical relationships 
among the units of information in XML documents.
\end{itemize}

The first aspect makes the new relevance definition much simpler than the current one, and more importantly, 
enables a straightforward integration in the unified relevance framework~\cite{Mizzaro}.
The second aspect
is directly inspired by the analysis of the level of agreement between the assessor and the users;
 indeed, the highest level of agreement was shown to be either on \emph{highly relevant} or on \emph{non-relevant} units of retrieval.
This means that both the assessor and users clearly agree upon the \emph{binary nature} of topical relevance of the retrieved units,
indicating that a unit is either \emph{relevant} or \emph{not} to an information need.
The second dimension of relevance, as introduced in the third aspect above, 
is completely orthogonal to the first dimension.
It is defined as follows.

The extent to which a unit of information is relevant to an information need is measured by considering the \emph{difference} between:
\begin{itemize}
\item The extent to which aspects of the information need are covered within the unit; and
\item The extent to which these aspects are covered within the other \emph{related} units (ancestors or descendants) in the document hierarchy.
\end{itemize}

For example, a relevant information unit is \emph{just right} to an information need if it mainly just covers aspects of the information need.
Alternatively, the information unit can be either \emph{too broad} or \emph{too narrow} to the information need.
A relevant information unit is \emph{too broad} if 
there is a \emph{descendant} that mainly just covers aspects of the information need.
Conversely, a relevant information unit is \emph{too narrow} if 
there is an \emph{ascendant} that is \emph{just right}.

The second dimension of relevance, as defined above, is very similar to \emph{document coverage} used in INEX 2002~\cite{Kazai-ECIR04}. 
Indeed, document (or component) coverage was used as a relevance dimension in INEX 2002 to measure how specific (or focused) the unit of retrieval is to 
the information need. List and de Vries~\cite{List-coverage} describe a formal approach to modelling the document coverage. 
Similar to our second dimension, some aspects of document coverage depend on the context where the information unit resides,
stating that ``the component is too small to act as a meaningful unit of information when retrieved by itself''~\cite{Kazai-ECIR04}.
This, however, makes the document coverage to also be dependent on the size of the retrieved unit.
The size of the unit of retrieval, on the other hand, is not explicitly considered in our relevance dimension.

\subsection*{New Relevance Definition for XML Retrieval}

Considering the above observations, we propose the following definition of relevance:

\begin{itemize}

\item An information unit is \emph{not relevant} to an information need if it does not cover any of the aspects of the information need; 

\item An information unit is \emph{relevant} to an information need if it covers any of the aspects of the information need. 
The extent to which the unit is relevant to the information need can be one of the following:
\begin{itemize}
\item \emph{Broad}, if the unit is too broad and includes other, non-relevant information; 
\item \emph{Narrow}, if the unit is too narrow and is part of a larger unit that better covers aspects of the information need; and
\item \emph{Just right}, if the unit mainly just covers aspects of the information need.
\end{itemize}

\end{itemize}

The above relevance definition has the following properties:

\begin{itemize}
\item In any one document path from the root element to a leaf, \emph{at most} one element can be \emph{Just right}.
However, multiple \emph{Just right} elements can exist in an XML document if they belong to different paths;
\item Every element in a path that resides \emph{above} the \emph{Just right} element is too broad, and only such elements are considered to 
be too broad; and
\item Every element considered to be too narrow is either a \emph{child} of an element that is \emph{Just right}, or a child of an element 
that is too narrow. Also, not every child of a relevant element has to be relevant.
\end{itemize}

There are two relevance dimensions described by the above definition: one based on the topical relevance, which uses a binary relevance scale
(\emph{relevant} or \emph{non-relevant}); 
and another based on hierarchical relationships among the information units in XML documents, which uses a three-graded relevance scale 
(\emph{Broad}, \emph{Narrow}, or \emph{Just right}).

\subsubsection*{Example Scenarios}

\begin{figure}[tb]
\epsfxsize=8.5cm
\centering\epsfbox{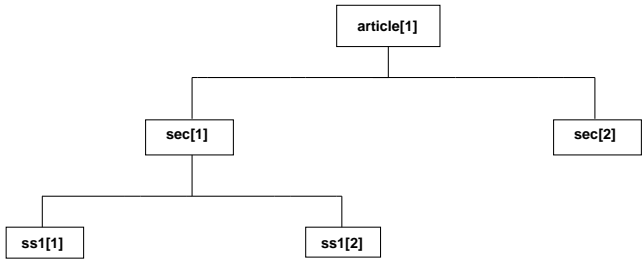}
\caption{A representation of an XML document}
\label{XML-tree}
\end{figure}

We further explain the new relevance definition with several example scenarios, with reference to the XML document representation in 
Figure~\ref{XML-tree}. 

\emph{Scenario 1}: Assume that only \verb+ss1[1]+ is relevant to an information need, 
and that it mainly just covers aspects of the information need. Because of the hierarchical relationships between the elements in the above 
document, both \verb+sec[1]+ and \verb+article[1]+ will also be relevant to the information need.
However, since \verb+ss1[2]+ contains no relevant information, \verb+sec[1]+ becomes \emph{too broad}. The same is also true for \verb+article[1]+. 
The set of relevant elements (or the full recall base) in this scenario consists of three elements:
one \emph{Just right} and two \emph{Broad}.

\emph{Scenario 2}: Assume that both \verb+ss1[1]+ and \verb+ss1[2]+ are relevant to an information need, 
and they also mainly just cover its aspects. 
The \verb+sec[1]+ element in this case contains two \emph{Just right} children, which also makes it \emph{Just right}. 
Indeed, the two \verb+ss1+ elements may cover two different aspects of the information need, or they may cover a single aspect 
from two different perspectives.
Since the additional context provided by \verb+sec[1]+ is (arguably) more desirable than each of the two separate contexts of its children,
both the \verb+ss1+ elements become \emph{too narrow}.
Also, since \verb+sec[2]+ contains no relevant information, \verb+article[1]+ becomes \emph{too broad}. 
The full recall base in this scenario consists of four relevant elements: one \emph{Just right}, one \emph{Broad}, and two \emph{Narrow}.

\emph{Scenario 3}: Assume that the three elements, \verb+ss1[1]+, \verb+ss1[2]+, and \verb+sec[2]+, are relevant to an information need, 
and all of them mainly just cover its aspects. 
The full recall base in this scenario consists of five relevant elements: one \emph{Just right} and four \emph{Narrow}, where \verb+article[1]+ 
is the only element that is \emph{Just right}.

\subsubsection*{Exploring Aspects of XML Retrieval}

Different aspects of XML retrieval may be explored by using the new relevance definition.

One aspect would be to measure the XML retrieval effectiveness when only \emph{Just right} elements are considered in the 
retrieval task. Note that in this case the full recall base consists of non-overlapping relevant elements, 
so there is no overlap problem during evaluation.

Another aspect would be to separately consider the \emph{Broad} and the \emph{Narrow} relevant elements in the recall base, 
and to measure the retrieval effectiveness against each of these elements.
Indeed, different topics (or queries) require different granularity or relevant elements~\cite{Pehcevski-Kluwer2005}.
However, in both of these cases different techniques may be needed to deal with the overlap problem. 

Previous work done by Voorhees in the field of Web retrieval
confirms the hypothesis that different retrieval techniques need be used to 
retrieve highly relevant, rather than just any relevant, Web pages~\cite{Voorhees-3-point}.
It may thus be worthwhile exploring whether, in the field of XML retrieval, different retrieval techniques would be needed
to retrieve \emph{Just right}, rather than any \emph{Broad} or \emph{Narrow}, relevant units of information.

\subsubsection*{Comparison with the INEX Relevance Definition}

Compared to the current INEX relevance definition, the new definition of relevance is much simpler. 
Indeed, instead of having a 10-point relevance scale that uses various combination of grades of the two INEX dimensions as values, 
the new relevance definition uses a four-point relevance scale with the following values: 
\emph{Non-relevant}, \emph{Narrow}, \emph{Just right}, and \emph{Broad}.

Also, more than one mappings may be possible between the INEX relevance definition and the new one.
For example, a partial mapping of the new four-point relevance scale to the INEX 10-point relevance scale is as follows.

\begin{enumerate}

\item \emph{Non-relevant} \verb+<=> E=0, S=0 (E0S0)+
 
\item \emph{Just right} \verb+<=> E=3, S=3 (E3S3)+

\item \emph{Broad} \verb+<=> E=3, S<3 (E3S2, E3S1)+

\item \emph{Narrow} \verb+<=> E<3, S=3 (E2S3, E1S3)+
\end{enumerate}

The above mapping is partial as it does not include the following four INEX relevance points: \verb+E2S2+, \verb+E2S1+, \verb+E1S2+, and \verb+E1S1+.
One reason for this is that we choose only a highly relevant (\verb+E3S3+) element on a path to represent a \emph{Just right} element.
From the properties of the new relevance definition (as outlined above), it follows that a \emph{Broad} or a \emph{Narrow} element 
could then be either \emph{above} or \emph{bellow} the \emph{Just right} element, which limits the mapping choices.
Another reason, however, stems from the fact that these four points of the relevance scale
were not well perceived by both the assessor and the users.
The latter may be the most probable cause for the observed inconsistencies regarding the \emph{Specificity} dimension.
Nevertheless, 
for the purposes of the evaluation of XML retrieval
there is almost no need to modify some of the current INEX metrics in order to use the new relevance definition.

The new relevance definition could also easily be applied to the recent proposal of performing 
the assessor's relevance judgements at INEX 2005. This proposal is as follows: 
first, for a returned article the assessor will be asked to highlight all of the relevant content.
Second, after the assessment tool automatically identifies the elements that enclose the highlighted content,
the assessor will need to judge the level of \emph{Exhaustivity} of these elements and of all their ancestors.
Last, based on the highlighted text, the level of \emph{Specificity} will be computed automatically as a ratio of relevant to non-relevant information, 
however a mapping may be needed to get the four relevance grades for the \emph{Specificity} dimension.

Although we agree that the above approach is very promising, it is still unclear 
whether keeping the current INEX relevance dimensions, along with their corresponding grades, 
would help reducing the cognitive load of the assessor (or the users) while performing the relevance judgements. 
The new relevance definition, on the other hand, is much simpler, and it also fits very nicely with the above proposal.

\section{Conclusions}

In this work, we have undertaken a detailed analysis of assessor's and users' behaviour in the context of XML retrieval.
We have shown that the two relevance dimensions used by INEX, \emph{Exhaustivity} and \emph{Specificity}, 
are not orthogonal and are perceived as one dimension by users.
By analysing the level of agreement between the assessor and the users,
we also wanted to identify how both of them perceive the points of the INEX 10-point relevance scale;
the results of our analysis show that the highest level of agreement 
is on the end points of the relevance scale, 
which means that a much simpler relevance scale would be a preferable choice for the field of XML retrieval.
We have proposed a new definition of relevance to be used by INEX,
and argued that its corresponding relevance scale is simpler and more comprehensive than the one currently used.

Our analysis also shows that, 
although the assessor handles the overlap problem differently than users,
in the users' case  
there is almost no overlap between the elements judged as relevant.
The latter confirms the hypothesis that users do not want to retrieve, and thus do not tolerate, redundant information. 

We have not discussed how the overlap problem may be modelled by the new relevance definition.
As argued previously, it may be possible to model the overlap problem by using a separate relevance dimension based on novel relevance, 
which can be integrated into the \emph{Context} component of the unified relevance framework~\cite{Mizzaro}.
However, in this paper we do not pursue this discussion any further.

The observed retrieval behaviour of the assessors and users was 
based on two topics, each from a different topic category. 
We did not observe any notable differences among the above behaviours for the two topics.
However, analysis of a greater number of topics is needed to confirm the significance of our findings.
This will enable a comparison between the observed and
the overall behaviour of the assessors and users, which will certainly establish the XML retrieval environment in a more consistent manner.
We leave the activities related to this analysis for future work.

It is our hope that, 
by analysing the different aspects of the observed retrieval behaviour,
the work presented in this paper will aid better understanding of the important issues surrounding INEX and 
the field of XML retrieval.

\section*{Acknowledgements}
We thank Saied Tahaghoghi and the anonymous reviewers for providing useful comments on earlier drafts of this paper.

\bibliographystyle{abbrv}
\bibliography{sigproc}  

\end{document}